\documentclass{article}

\usepackage{amsmath}
\usepackage{comment}
\usepackage{graphicx}
\usepackage{cite}
\usepackage{indentfirst}
\usepackage{xcolor}
\usepackage[margin=0.8in]{geometry}
\usepackage{multicol}

\begin{document}

\section*{Transversal transport of magnons in a modified Lieb lattice}

\noindent P. G. de Oliveira$^a$* \newline
\noindent E-mail address: pgopedro@ufmg.br  \newline

\noindent and \newline

\noindent A. S. T. Pires$^a$ \newline
\noindent E-mail address: antpires@fisica.ufmg.br
\vspace{\baselineskip}

\noindent $^a$ Departamento de Física, Universidade Federal de Minas Gerais, Belo Horizonte, MG, CP702, 30123-970, Brazil.
\noindent * Corresponding author.

\subsection*{Abstract}

\noindent We studied a two-band magnon insulating model whose geometry is that of a modified Lieb lattice \textcolor{black}{in} which one of the sites was removed. \textcolor{black}{Anisotropic ferromagnetic exchange interactions exist between the three nearest neighbors, and the anisotropy opens a gap in the magnon energy band structure}. A non-vanishing Berry curvature is induced by a Dzyaloshinskii-Moriya interaction (DMI). The topology of the bands is trivial (in the sense of a null Chern number), but the finite Berry curvature induces Hall-like transport effects whose coefficients were calculated. \textcolor{black}{Their dependence on temperature was studied and shows a resemblance with other magnon insulating systems found in the literature. The dependence on exchange couplings, DMI parameter, and external magnetic field was also investigated.}

\vspace{\baselineskip}

\noindent Keywords: Spin waves; magnons; Dzyaloshinskii–Moriya interaction; transport; Hall-like effects.

\newpage
\section{Introduction}

Topological effects \textcolor{black}{in} condensed matter systems have been intensely studied since the discovery of the quantum Hall effect by von Klitzing \textit{et al.} \cite{klitzing}. The attention falls naturally on the so-called topological insulators (TIs),  \textcolor{black}{electronic systems with gapped bands in the bulk,} and robust conducting edge (or surface) modes. These systems have different insulating phases characterized by topological indices \cite{topological} and may show Hall-like effects when subjected to a field or temperature gradient \cite{she,intrinsic_she,she_science,ahe,prl97}. These effects arise in materials with strong spin-orbit coupling \textcolor{black}{and can be related to the electronic bands' Berry phase and Berry curvature \cite{reviewberryphase}}.

In analogy to TIs in electronic systems, \textcolor{black}{topological magnon insulators (TMIs) are gapped magnonic systems which present non-trivial topology.} Magnons are spin-wave excitations of the ground state of \textcolor{black}{localized spin systems. When magnon bands have finite Berry curvature,} the same Hall-like transport effects can arise \cite{halleffect3,halleffect4,halleffect5,thermal,prb89,LeeHanLee,nthermal,Meier2003,PRL103}. Since magnons are bosons, magnonic systems are intrinsically different from electronic ones, which motivates their study. \textcolor{black}{A notable fact is that magnons favor dissipationless transport because of their uncharged nature, which is of great interest to spintronics \cite{spintronics}}. While topological effects in magnon systems were first discovered in a three-dimensional material with the geometry of the pyrochlore lattice \cite{science_onose}, the main theoretical interest nowadays falls on two-dimensional lattices, where the most studied geometries are the honeycomb lattice \cite{owerre2016,owerre2,honeycomb2} and the kagome lattice \cite{thermal,kagome_mook2014,kagomeAFM,kovalev2016,LeeHanLee}. 
The latter can be layered with triangular planes to form the pyrochlore structure. Other lattices that were predicted to hold topological magnon effects are the Shastry-Sutherland \cite{ss}, square \cite{square}, checkerboard \cite{checkerboard1,checkerboard2,pires2020} and Lieb \cite{lieb} lattices.

\begin{figure}[h!]
\centering
\includegraphics[width=0.4\textwidth]{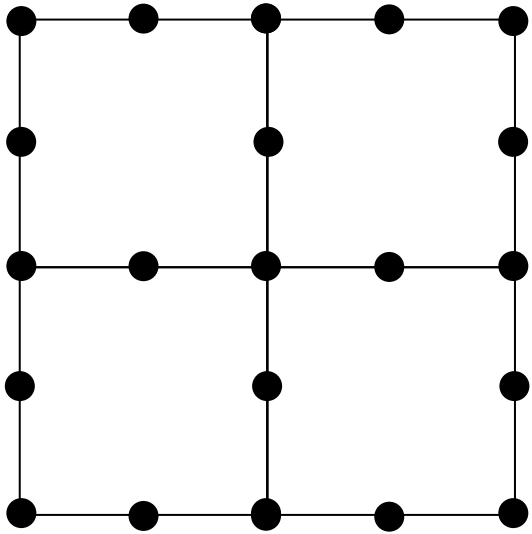}
\caption{The Lieb lattice.}
\label{fig0}
\end{figure}

\textcolor{black}{The Lieb lattice (Figure \ref{fig0}) is particularly interesting because it is the geometry that $CuO_2$ planes assume in high-$T_c$ cuprate superconductors \cite{lieb_cuprates}. It has also been identified in some organic compounds as a ``hidden lattice'' \cite{org}.
In a tight-binding approach, the Lieb lattice is a three-band model that shows a flat band \cite{prb87} and a single Dirac cone in the Brillouin zone. The energy gap can be opened by creating a TI phase by an intrinsic spin-orbit interaction term \cite{prb82_weeks,prb86}. The Hubbard model in the Lieb lattice was studied in Ref. \cite{Hubbard} using Monte Carlo simulations and revealed a phase change between a metallic and an Anderson insulator phase. A Hartree-Fock approach for interacting electrons was used in Ref. \cite{HF}, showing many topological phases for the Lieb lattice.
The Berry curvature and anomalous Hall effect of the electronic Lieb lattice were studied by He \textit{et al.} \cite{prb85_he2012}.}

\textcolor{black}{Magnonic systems in the Lieb lattice have also been intensely investigated in the context of the Heisenberg model. In Ref. \cite{lieb} it was shown that a complex hopping between next-near-neighbours can induce topological insulating phases. The insulating system presents thermal spin Hall effect, and the temperature dependence of the thermal Hall conductivity is related to its topological phases. Linear response theory and the Green function approach were used to investigate Heisenberg-like models in the Lieb lattice by Yarmohammadi \textit{et al.} \cite{Y12,Y10,Y13,Y8,Y2,Y4,Y5,Y7} within the magnon picture. Several properties of the system were investigated, like density of magnon modes, magnetic susceptibility, magnon heat capacity \cite{Y12}, optical absorption \cite{Y10}, dynamical thermal conductivity \cite{Y13}, and magneto-topological property \cite{Y8}. The effect of charged impurities in the Lieb lattice was studied in Refs. \cite{Y2,Y4,Y5,Y7}.}

\textcolor{black}{Based on the necessity of finding novel magnetic systems with non-trivial topologies, we propose a study of a modified version of the ferromagnetic Lieb lattice, where one of the inequivalent sites is removed. The topological effects are induced by a Dzyaloshinskii-Moriya interaction (DMI) between the next-next-near neighbors. The DMI breaks the time-reversal symmetry (TRS), and is the most common way of inducing non-vanishing Berry curvature and topological effects.} We present the system's geometry and Hamiltonian in section \ref{model}, calculate its magnon band structure in section \ref{bandstructure}, calculate and discuss its Hall-like transport coefficients in section \ref{t_coeff} and present our conclusions in section \ref{conclusion}.

\section{Model}
\label{model}

We consider a lattice with two inequivalent sites (A and B) in each square
unit cell, with the following Hamiltonian:

\bigskip%

\begin{align}
H    =&-J_{1}\sum_{\left\langle i,j\right\rangle }\left(  S_{i}^{x}S_{j}%
^{x}+S_{i}^{y}S_{j}^{y}+\lambda S_{i}^{z}S_{j}^{z}\right)  -J_{2}%
\sum_{\left\langle \left\langle i,j\right\rangle \right\rangle \in A}\left(
S_{i}^{x}S_{j}^{x}+S_{i}^{y}S_{j}^{y}+\lambda S_{i}^{z}S_{j}^{z}\right)
\nonumber\\
&  -J_{3}\sum_{\left\langle \left\langle \left\langle i,j\right\rangle
\right\rangle \right\rangle }\left(  S_{i}^{x}S_{j}^{x}+S_{i}^{y}S_{j}%
^{y}+\lambda S_{i}^{z}S_{j}^{z}\right) \nonumber\\
&  -D\sum_{\left\langle \left\langle \left\langle i,j\right\rangle
\right\rangle \right\rangle } \nu_{ij}\left(  S_{i}^{x}S_{j}^{y}-S_{i}^{y}%
S_{j}^{x}\right)  -B\sum_{i}S_{i}^{z} \label{hamiltonian}%
\end{align}

\bigskip

\begin{figure}[h!]
\centering
\includegraphics[width=0.4\textwidth]{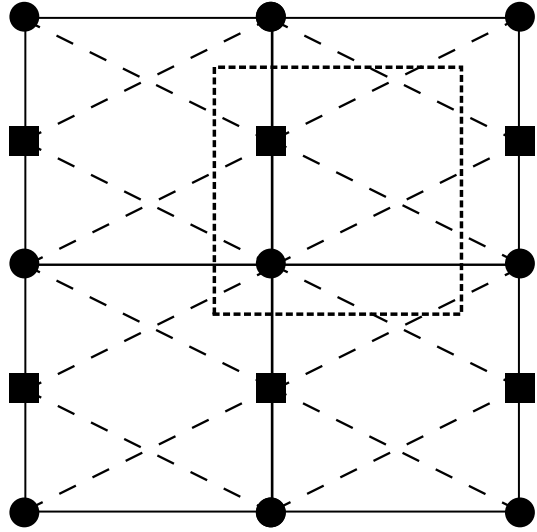}
\caption{Modified Lieb lattice studied \textcolor{black}{in} this paper. Circles (squares) represent the A (B) sites. Solid lines are ferromagnetic exchange bonds ($J_1$ between A and B; $J_2$ between two A sites). Dashed lines
represent both $J_3$ and DM interaction. \textcolor{black}{The dashed square represents the unit cell.}}
\label{rede}
\end{figure}

The lattice can be seen as a modified Lieb lattice in
which one of the three inequivalent sites was removed. The unit cell is a square \textcolor{black}{whose side length was set as the unit}. A sketch of the lattice
can be seen in Figure \ref{rede}. There are ferromagnetic exchange interactions
(solid lines) between near-neighbors $A$ and $B$ (strength $J_{1}$) and
next-near-neighbors $A$ (strength $J_{2}$). On the diagonals between $A$ and
$B$ (dashed lines), there are a $J_{3}$ exchange
interaction and a Dzyaloshinskii-Moriya interaction\ (DMI) \cite{dz,moriya} of the form
$-\mathbf{D}_{ij}\cdot\left(  \mathbf{S}_{i}\times\mathbf{S}_{j}\right)  $. The
\textcolor{black}{latter} is responsible for the finite Berry curvature and Hall-like transport effects,
which are the focus of this paper.

\textcolor{black}{Moriya's rules do not forbid the DM interaction}, for there is no center of inversion
at the midpoint of the bond \cite{moriya}. Hence we can consider the interaction as a regular
DMI, and there is no need \textcolor{black}{to introduce an external electric field to induce
the interaction, as is the case} in some lattices where the regular DMI is
forbidden \cite{externalDM}. The same Moriya's rules restrict the DM vector to the $z$ direction, as a two-dimensional lattice is symmetric \textcolor{black}{with respect to a reflection upon its plane}. We take $\mathbf{D}_{ij}=D \nu_{ij}\mathbf{\hat{z}}$ (fourth term in (\ref{hamiltonian})), where $\nu_{ij}=\pm1$ for
different bond directions, following Figure \ref{DM}. \textcolor{black}{In all exchange interactions,} there is an anisotropy $\lambda>1$ on the $z$ direction, which is responsible for the stabilization of the magnetic order in an easy-axis configuration \cite{Herring1951,Bruno1991} (otherwise, the system would have no long-range order according to the Mermim-Wagner theorem). The last term
in (\ref{hamiltonian}) is a Zeeman interaction with a constant magnetic field
$\mathbf{B=}B\mathbf{\hat{z}.}$

\bigskip

\begin{figure}[h!]
\centering
\includegraphics[width=0.4\textwidth]{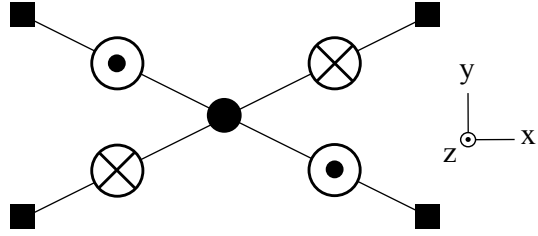}
\caption{Configuration of the DMI vectors $\textbf{D}$ on the diagonal bonds, departing from an $A$ site.}
\label{DM}
\end{figure}

\section{Magnon bands}
\label{magnonbands}
We are interested in the linear spin-wave regime, so we use the zeroth order expansion of the Holstein-Primakoff representation:

\begin{align}
S_{i}^{+}  &  \approx  \sqrt{2S} \ a_{i}\text{ ; }S_{i}^{-} \approx \sqrt{2S} \ a_{i}^{\dagger}\text{ ; }%
S_{i}^{z}=S-a_{i}^{\dagger}a_{i}\\
S_{j}^{+}  &  \approx \sqrt{2S} \ b_{j}\text{ ; }S_{j}^{-} \approx \sqrt{2S} \ b_{j}^{\dagger}\text{ ; }%
S_{j}^{z}=S-b_{j}^{\dagger}b_{j}\nonumber
\end{align}

We are studying a ferromagnetic model, but since we have two inequivalent sites, we
need two operators, $a$ and $b$. \textcolor{black}{The index $i$ represents sites in sublattice $A$} and $j$ in sublattice $B$. This spin-wave approximation works for low \textcolor{black}{temperatures}, and the magnon-magnon interactions (higher order \textcolor{black}{terms} on the expansion) are much smaller than the linear contribution and can be neglected \cite{Dyson1956,Oguchi1960}.

After transforming, keeping only the quadratic terms and applying a Fourier
transform, we can write the harmonic momentum-space Hamiltonian as:

\bigskip%

\begin{equation}
H_{harm}=S\sum_{k}\psi_{k}^{\dagger}\left(  h_{0}\hat{1}_{2}+\hat{M}_k\right)
\psi_{k}%
\label{hamilt_inicial}
\end{equation}

where $\psi_{k}^{\dagger}=\left(  a_{k}^{\dagger}\text{ \ \ }b_{k}^{\dagger}\right)  .$
Here, $h_{0}$ and \textcolor{black}{the elements of the} $2\times2$ matrix $\hat{M}_{k}$ are

\begin{align}
h_{0}  &  =J_{2}\left(  \lambda-\cos k_{x}\right)  +2\lambda\left(
J_{1}+2J_{3}\right)  +\frac{B}{S}\\
M_{11}  &  =J_{2}\left(  \lambda-\cos k_{x}\right) \nonumber\\
M_{12}  &  =M_{21}^{\ast}=-2J_{1}\cos\frac{k_{y}}{2}-4J_{3}\cos k_{x}%
\cos \frac{k_{y}}{2}  -4iDm_{k}\nonumber\\
M_{22}  &  =-J_{2}\left(  \lambda-\cos k_{x}\right) \nonumber
\end{align}

where $m_{k}= -\sin k_{x}\sin  \frac{k_{y}}{2}$. \textcolor{black}{All the calculations in this study were made for a lattice with an infinite number of sites with periodic boundary conditions, so the wave vectors \textbf{k} are not quantized and continuously cover the Brillouin zone ($k_x=[-\pi,\pi]$, $k_y=[-\pi,\pi$]). }
We can write the matrix $\hat{M}_{k}$ as a Pauli vector

\bigskip%

\begin{equation}
\hat{M}_k=h_{x}\left(  \mathbf{k}\right)  \hat{\sigma}_{x}+h_{y}\left(
\mathbf{k}\right)  \hat{\sigma}_{y}+h_{x}\left(  \mathbf{k}\right)
\hat{\sigma}_{z}%
\label{pauli}
\end{equation}

\bigskip

\textcolor{black}{so that the dispersion relation is \cite{book_pires}
\begin{equation}
\frac{E_{\pm}}{\hbar}=\omega_{\pm}\left(  \mathbf{k}\right)  =S\left(  h_{0}\left(
\mathbf{k}\right)  \pm h\left(  \mathbf{k}\right)  \right)
\end{equation}}

\textcolor{black}{while the eigenstates are
\begin{equation}
u_{+}(\textbf{k})=
\begin{pmatrix}
p_1e^{-i\theta} \\
p_2 
\end{pmatrix}, \ \ \ \ \ u_{-}(\textbf{k})=
\begin{pmatrix}
p_2e^{-i\theta} \\
-p_1 
\end{pmatrix}
\label{eigenstates}
\end{equation}  
}

\textcolor{black}{Here we used the definitions}

\textcolor{black}{
\begin{equation}
h\left(  \mathbf{k}\right)  = \left\| \mathbf{h} \left(  \mathbf{k}\right) \right\| =\sqrt{h_{x}^{2}\left(  \mathbf{k}\right)
+h_{y}^{2}\left(  \mathbf{k}\right)  +h_{z}^{2}\left(  \mathbf{k}\right)  }
\end{equation}
\begin{equation}
p_1=\sqrt{\frac{h+h_z}{2h}}
\end{equation}
\begin{equation}
p_2=\sqrt{\frac{h-h_z}{2h}}
\end{equation}
\begin{equation}
tan \ \theta=\frac{h_y}{h_x}
\end{equation}
}

\textcolor{black}{We can see that the parameters $h_0$, $h_x$, $h_y$ and $h_z$ determine ultimately the eigenvalues and eigenstates of the system. The expressions above can be applied to any Hamiltonian that can be written in the form of Eq. \ref{hamilt_inicial}. For more details, see Ref. \cite{book_pires}}.

\textcolor{black}{For the model studied here, the parameters are:}

\begin{align}
h_{x}\left(  \mathbf{k}\right)   &  =-2\cos  \frac{k_{y}}{2}
\left(  J_{1}+2J_{3}\cos k_{x}\right) \\
h_{y}\left(  \mathbf{k}\right)   &  =4Dm_{k}\nonumber\\
h_{z}\left(  \mathbf{k}\right)   &  =J_{2}\left(  \lambda-\cos k_{x}\right)
\nonumber
\end{align}

Sometimes it is useful to perform the substitution $\tan\phi=\frac{D}{J_{3}},$
arriving at%

\begin{equation}
h_{x}\left(  \mathbf{k}\right)  =-2J_{1}\cos  \frac{k_{y}}{2}
-4J_{D}\gamma_{k}\cos\phi\text{ \ , \ }h_{y}\left(  \mathbf{k}\right)
=4J_{D}m_{k}\sin\phi
\end{equation}

where $\gamma_{k}=\cos k_{x}\cos  \frac{k_{y}}{2}$ and $J_{D}=\sqrt
{J_{3}^{2}+D^{2}}$. The phase $\phi$ can be seen as a magnetic flux generated
by the DM term. In the case of a pure DM interaction in the diagonals (no
exchange $J_{3}$),\ the phase is $\phi=\pi/2$.

The explicit dispersion relation is

\begin{align}
\omega_{\pm}\left(  \mathbf{k}\right)   &  =SJ_{2}\left(  \lambda-\cos
k_{x}\right)  +2S\lambda\left(  J_{1}+2J_{3}\right) + B +  \label{eqdispersion} \\
&  \pm S\sqrt{4\cos^{2} \frac{k_{y}}{2}  \left(  J_{1}%
+2J_{3}\cos k_{x}\right)  ^{2}+J_{2}^{2}\left(  \lambda-\cos k_{x}\right)
^{2}+16D^{2}m_{k}^{2}} \nonumber
\end{align}

\begin{figure}[h!]
\centering
\includegraphics[width=0.5\textwidth]{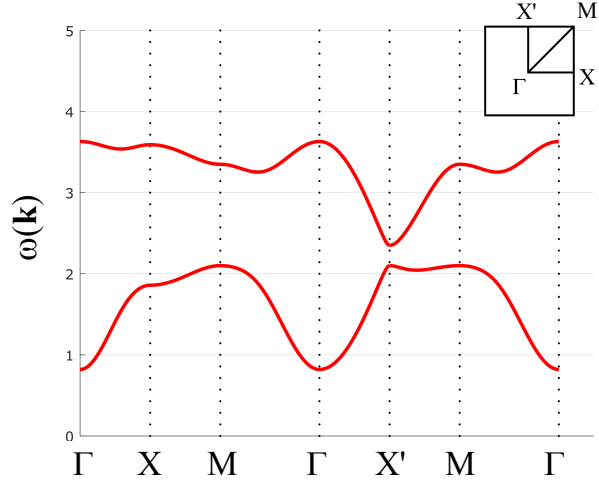}
\caption{Band structure of \textcolor{black}{the} system for $S=1/2$, $B=0$, $J_1=1$, $J_2=0.5$,
$J_3=0.2$, $D=0.1$ and $\lambda=1.5$.}
\label{bandstructure}
\end{figure}

The band structure of the system is plotted in Figure \ref{bandstructure} for  $S=1/2$, $B=0$, $J_1=1$, $J_2=0.5$,
$J_3=0.2$, $D=0.1$ and $\lambda=1.5$. There is a gap of $2SJ_{2}\left(  \lambda-1\right)$ at the high-symmetry point $X^{\prime}=\left(  0,\pm\pi\right) $. The gap
vanishes into a Dirac point in the isotropic limit ($\lambda=1$) independently of the value of
other parameters. Therefore, the anisotropy is responsible for the gap, \textcolor{black}{not the DM interaction, as with other magnon-insulating systems} 
\cite{owerre2016,honeycomb2,lieb,pires2020}. For a zero applied field $B=0$ and small enough values of $D$ the \textcolor{black}{lower} band has a single global minimum at the $\Gamma$ point, which has zero value if and only if $\lambda=1$. Hence, \textcolor{black}{we have a gapless system with a Goldstone mode in the isotropic limit}. For our purposes, this possibility is absent since we defined $\lambda > 1$ for an easy axis configuration. Therefore, we have an insulating system with no Goldstone mode.

\begin{figure}[h!]
\centering
\includegraphics[width=0.5\textwidth]{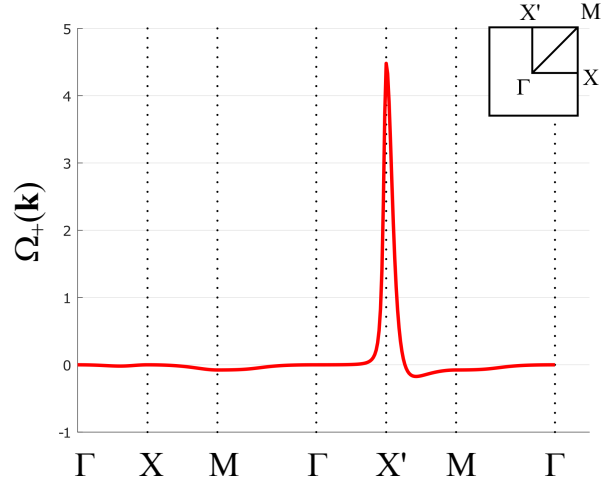}
\caption{Berry curvature of the \textcolor{black}{upper} band of the system for $S=1/2$, $J_1=1$, $J_2=0.5$,
$J_3=0.2$, $D=0.1$ and $\lambda=1.5$.}
\label{berry}
\end{figure}

\section{Transport coefficients}
\label{t_coeff}

\bigskip

For magnon systems, it is known that an external in-plane magnetic field
gradient generates not only a parallel spin current but also a spin response
\textcolor{black}{in} the transverse direction \cite{halleffect3}:

\bigskip

\begin{equation}
j_{x}^{S}=-\sigma_{xy}\left(  \partial_{y}B\right)
\end{equation}
\bigskip

\textcolor{black}{That} is the so-called \textit{spin Hall effect} of magnons, and it is
responsible for a protected spin current on the edges of a 2D magnet
\cite{Meier2003,PRL103}. In a semiclassical picture, this current can be
explained by the effect of the borders, which act as an effective field
gradient, confining the magnons inside the magnet and generating a spin
current perpendicular to the gradient (along the edge)$\cdot$

The transverse conductivity $\sigma_{xy}$ can be obtained from the Berry
curvature of the system \cite{halleffect4,halleffect5}:

\bigskip

\begin{equation}
\sigma_{xy}=-\frac{1}{\hbar V_{BZ}}\sum_{\lambda}\int_{BZ}dk_{x}dk_{y}\text{
}n_{\lambda}\left(  \mathbf{k}\right)  \Omega_{\lambda}\left(  \mathbf{k}%
\right)
\end{equation}

\bigskip

where $\lambda$ sweeps the magnon energy bands. Here, $n_{\lambda}\left(
\mathbf{k}\right)  =\left(  e^{\hbar\omega_{\lambda}\left(  \mathbf{k}\right)
/k_{B}T}-1\right)  ^{-1}$ is the Bose-Einstein distribution and $\Omega
_{\lambda}\left(  \mathbf{k}\right)  $ is the off-plane component of the (vector) Berry curvature of the band defined as \cite{halleffect4,halleffect5,reviewberryphase}:

\bigskip%

\begin{equation}
\mathbf{\Omega_{\lambda}}\left(  \mathbf{k}\right)  =i\left\langle \nabla_{\mathbf{k}%
}u_{\lambda}\left(  \mathbf{k}\right)  \right\vert \times\left\vert
\nabla_{\mathbf{k}}u_{\lambda}\left(  \mathbf{k}\right)  \right\rangle
\end{equation}

\bigskip

\textcolor{black}{where $\left\vert u_{\lambda}\left(  \mathbf{k}\right)  \right\rangle $ is the
Bloch wave function (eigenstate) of the $\lambda$ band. Using the analytical expression of the eigenstates (see Eq. \ref{eigenstates}), the off-plane component of the Berry curvatures can be written as \cite{honeycomb2,reviewberryphase,pires2020}}

\bigskip

\begin{equation}
\Omega_{+}\left(  \mathbf{k}\right)  =-\frac{1}{2h^{3}}\mathbf{h} \cdot 
\left(  \partial_{k_{x}}\mathbf{h} \times \partial_{k_{y}}\mathbf{h}\right)
\end{equation}

\bigskip

and $\Omega_{-}\left(  \mathbf{k}\right)  =-\Omega_{+}\left(
\mathbf{k}\right)  $. For our system the definition above gives:

\begin{align}
\Omega_{+}\left(  \mathbf{k}\right)  =& -2\frac{DJ_{2}}{h\left(  \mathbf{k}\right)^3} \Bigg\{ J_{1}\left[  \sin^{2}%
k_{x}-\left(  \lambda-\cos k_{x}\right)  \cos k_{x}\sin^{2}  \frac
{k_{y}}{2}  \right]  + \label{berry_eq} \\
&+2J_{3}\left[  \left(  \lambda-\cos k_{x}\right)
\left(  \sin^{2}k_{x}\cos^{2} \frac{k_{y}}{2}  -\cos^{2}%
k_{x}\sin^{2} \frac{k_{y}}{2}  \right)  +\sin^{2}k_{x}\cos
k_{x}\right]  \Bigg\}  \nonumber
\end{align}

The Berry curvature of the \textcolor{black}{upper} band through the Brillouin zone is plotted
in Figure \ref{berry}. As \textcolor{black}{expected}, the curvature \textcolor{black}{is concentrated mainly
around} the point in the Brillouin zone where the energy gap occurs \cite{reviewberryphase}.

The Chern number is defined as proportional to the integral of the Berry curvature in the
Brillouin zone. \textcolor{black}{It is} an integer number \textcolor{black}{that} labels the inequivalent
topological phases in Chern insulators \cite{topological}. \textcolor{black}{Despite} the finite Berry curvature, the Chern numbers of the bands are null for any combination of parameter values (provided that $\lambda > 0$), which
means that the system doesn't present a non-trivial topological insulating phase. Nevertheless,
the non-vanishing Berry curvature gives rise to Hall-like effects like the spin Hall effect shown above.

The Berry curvature is related to transverse spin and heat currents in
response to an applied temperature gradient \cite{nthermal}. These
are the \textit{spin Nernst effect} \cite{nernst},

\bigskip

\begin{equation}
j_{x}^{N}=-\alpha_{xy}\left(  \partial_{y}T\right)
\end{equation}

\bigskip

and the \textit{thermal Hall effect} \cite{halleffect4,halleffect5,thermal}

\bigskip

\begin{equation}
j_{x}^{Q}=-\kappa_{xy}\left(  \partial_{y}T\right)  .
\end{equation}

\bigskip

Here, $j_{x}^{N}$ is the spin current and $j_{x}^{Q}\,$\ the heat
current. The transport coefficients are defined as \cite{halleffect4,halleffect5,nernst,owerre2}:

\begin{align}
\alpha_{xy}  & =-\frac{k_{B}}{\hbar V_{BZ}}\sum_{\lambda}\int_{BZ}dk_{x}%
dk_{y}\text{ \ }c_{1}\left(  n_{\lambda}\left(  \mathbf{k}%
\right)\right)   \Omega_{\lambda}\left( \mathbf{k}\right) \nonumber\\
\kappa_{xy}  & =-\frac{k_{B}^{2}T}{\hbar V_{BZ}}\sum_{\lambda}\int_{BZ}%
dk_{x}dk_{y}\text{ \ }c_{2}\left(  n_{\lambda}\left(  \mathbf{k}%
\right)\right)  \Omega_{\lambda}\left(  \mathbf{k}%
\right)
\end{align}

with

\begin{align}
c_{1}\left(  x\right)   &  =\left(  1+x\right)  \ln\left(  1+x\right)  -x\ln
x \nonumber\\
c_{2}\left(  x\right)   &  =\left(  1+x\right)  \ln\left(  \frac{1+x}%
{x}\right)  ^{2}-\left(  \ln x\right)  ^{2}-2Li_{2}\left(  -x\right)
\end{align}

where $Li_{2}\left(  x\right)  $ is Spence's dilogarithm function.

\begin{figure}[h!]
\centering
\includegraphics[width=0.5\textwidth]{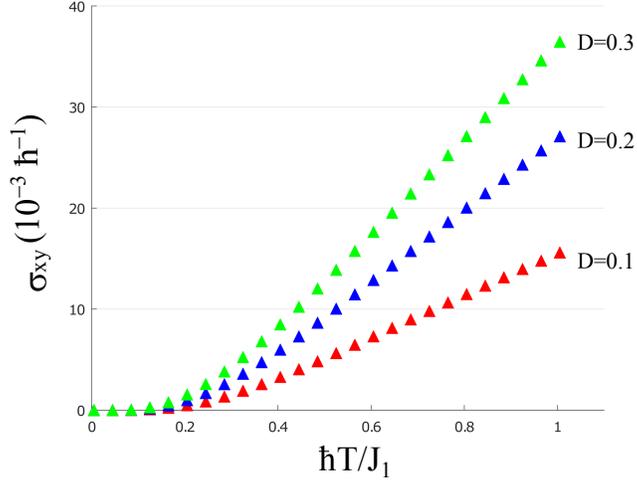}
\caption{Spin Hall conductivity versus temperature for $S=1/2$, $B=0$, $J_1=1$, $J_2=0.5$, $J_3=0$, $\lambda=1.2$ and different values of $D$.}
\label{cond}
\end{figure}

The spin Hall conductivity $\sigma_{xy}$ is plotted as a function of temperature for three values of $D$ in Figure \ref{cond} (in all transport coefficients plots the temperature and applied magnetic field are shown in units of the exchange energy $J_1$). The plot shows a monotonically rising behavior of $\sigma_{xy}$, similar to what could be observed in the checkerboard lattice \cite{pires2020,checkerboard2}. At zero temperature $\sigma_{xy}$ is zero due to the absence of magnon excitations. That is a consequence of the fact that boson numbers are not conserved and vanish in the zero temperature limit. \textcolor{black}{However, magnons are thermally excited as the temperature increases from zero, and} $\sigma_{xy}$ becomes finite. At low temperatures, the lower band dominates.

In all temperature plots, the range of the temperature axis was chosen to show the character of the curve \textcolor{black}{best}, but we wouldn't expect the model to work at such high temperatures. \textcolor{black}{We should remember that} the linear spin-wave approximation deals with perturbations of the ordered ground state and only works at low temperatures. \textcolor{black}{One} way to quantify the validity of the model is \textcolor{black}{by} calculating the expected value of the total boson number:

\begin{equation}
\Delta=\frac{1}{V_{BZ}}\sum_{\lambda}\int_{BZ}dk_{x}dk_{y} \ n_{\lambda}
\end{equation}

The spin-wave approach works for $\Delta \ll S$. Some values of $\Delta$ for different spins and temperatures are represented in Table \ref{table}. As we can see, the approximation works better for low temperatures and high values of spin.

\begin{table}[h!]
\centering
\includegraphics[width=0.5\textwidth]{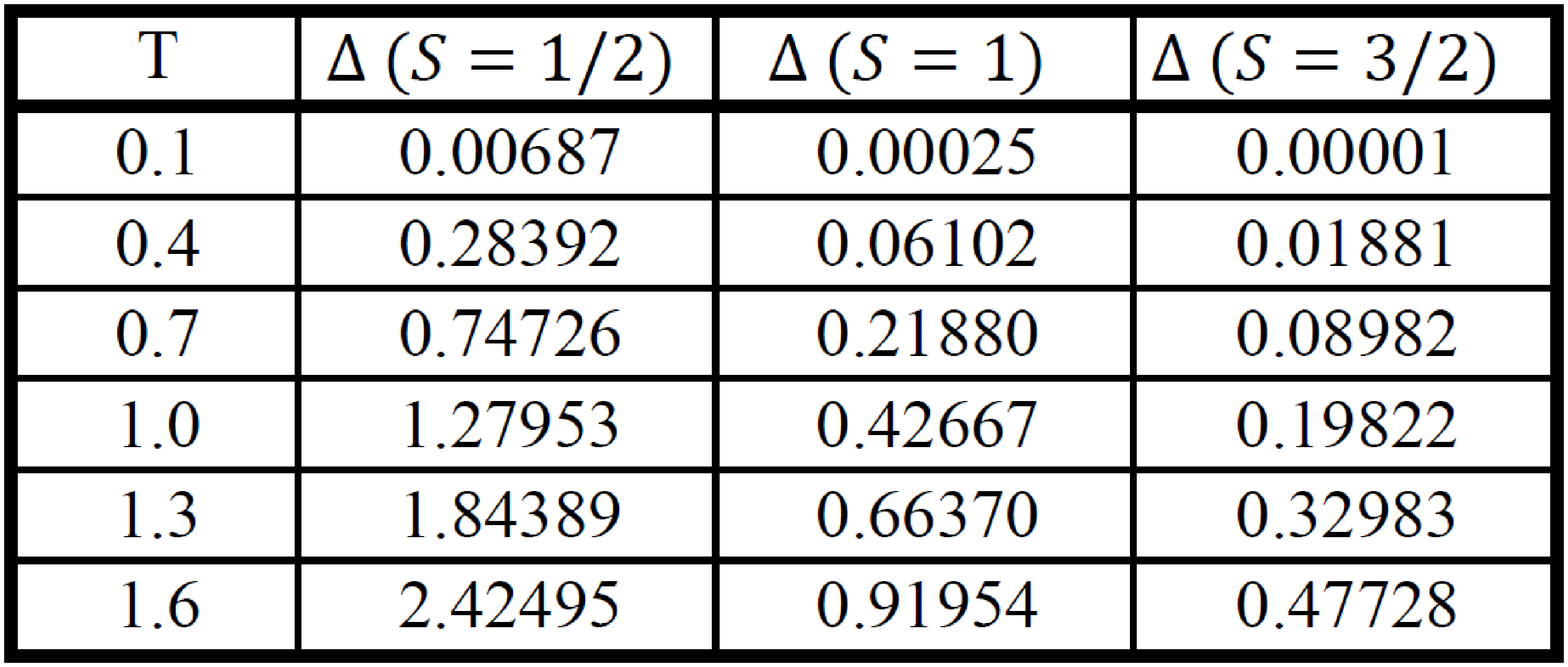}
\caption{Expectation value of the boson number for different temperatures and \textcolor{black}{spins}. The theory parameters are $B=0$, $J_1=1$, $J_2=0.5$, $J_3=0$, $\lambda=1.2$, and $D=0.2$. \textcolor{black}{The temperature} is in units of $J_1$.}
\label{table}
\end{table}

\begin{figure}[h!]
\centering
\includegraphics[width=0.5\textwidth]{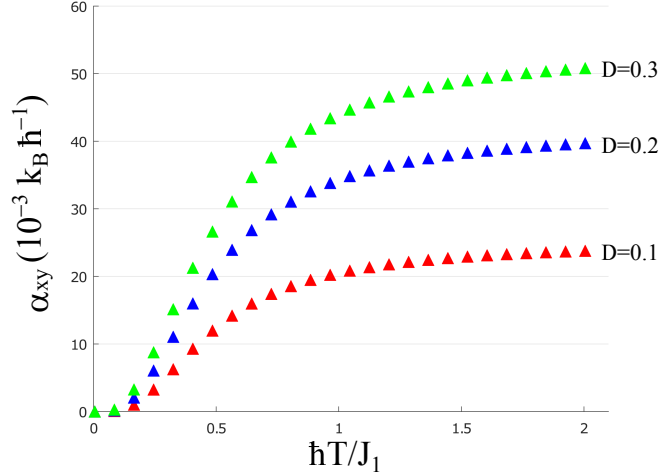}
\caption{Spin Nernst coefficient versus temperature for $S=1/2$, $B=0$, $J_1=1$, $J_2=0.5$, $J_3=0$, $\lambda=1.2$ and different values of $D$.}
\label{sn}
\end{figure}

In Figure \ref{sn}, we present the spin Nernst coefficient as a function of temperature. The coefficient $c_1(x)$ decreases with $x$, \textcolor{black}{leading} to a flattening of $\alpha_{xy}$ for high temperature. We see a monotonic response to the temperature, similar to other magnon systems like the AFM checkerboard and FM Kagome lattices \cite{checkerboard1,kovalev2016}, but in contrast to the FM checkerboard and both FM and AFM honeycomb lattices \cite{pires2020,honeycomb2,nernst}.

\begin{figure}[h!]
\centering
\includegraphics[width=0.5\textwidth]{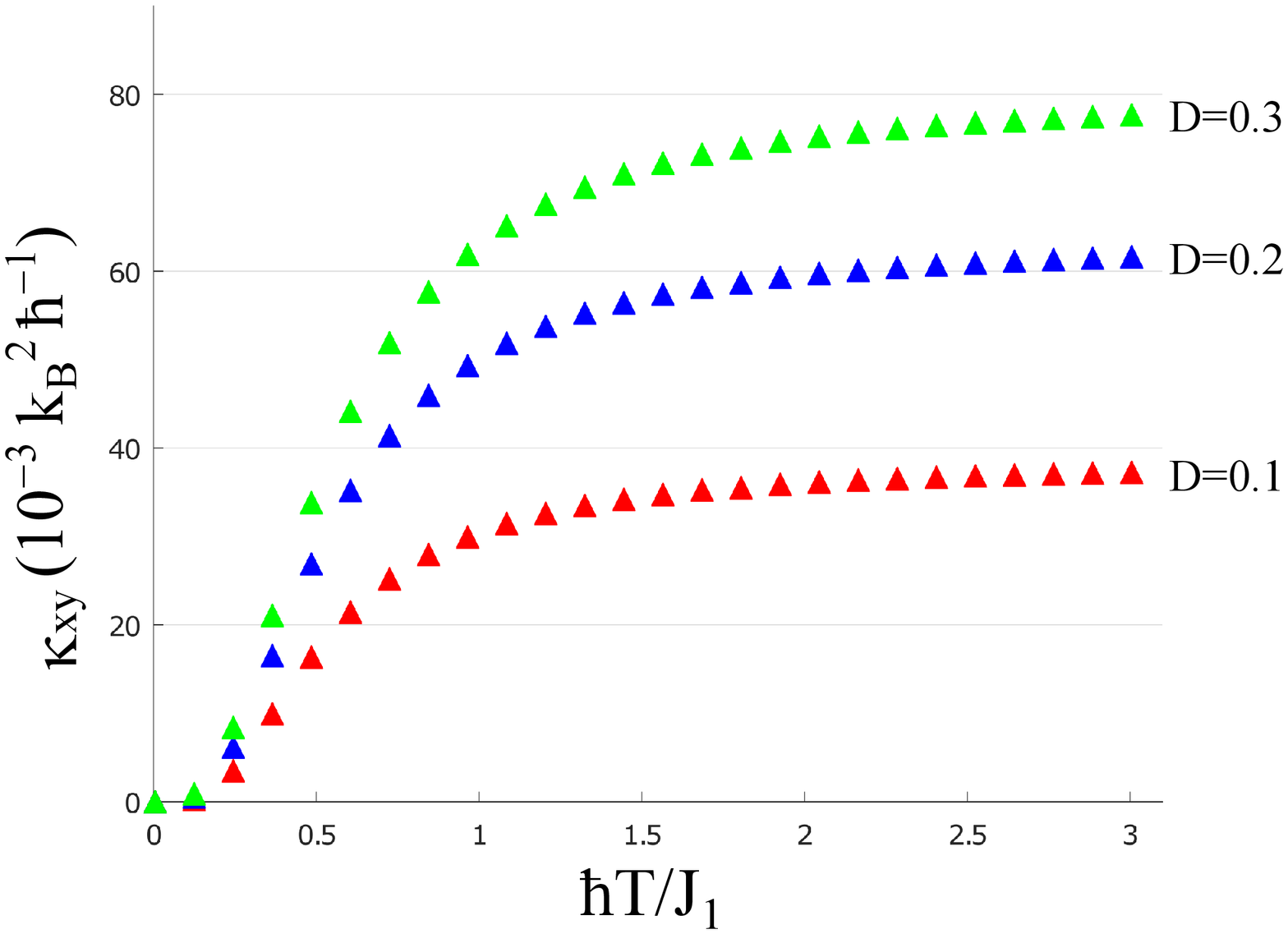}
\caption{Thermal Hall conductivity versus temperature for $S=1/2$, $B=0$, $J_1=1$, $J_2=0.5$, $J_3=0$, $\lambda=1.2$ and different values of $D$.}
\label{th}
\end{figure}

\begin{figure}[h!]
\centering
\includegraphics[width=0.5\textwidth]{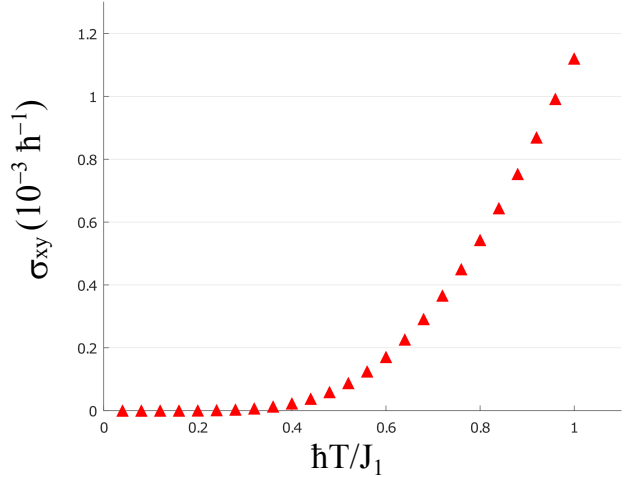}
\caption{Spin Hall conductivity versus temperature for $S=1/2$, $B=0$, $J_1=J_2=J_3=1$, $D=0.2$ and $\lambda=1.2$. \textcolor{black}{This plot exemplifies that the transport coefficient curves change only quantitatively for different combinations of parameters}.}
\label{2D}
\end{figure}

The thermal Hall conductivity versus temperature is shown in Figure \ref{th} \textcolor{black}{and increases monotonically} with no sign change. This \textcolor{black}{behavior} is similar to the FM honeycomb, AFM checkerboard and AFM Kagome lattices \cite{owerre2,checkerboard2,kagomeAFM}. The FM Lieb and Kagome lattices also show the same \textcolor{black}{behavior} for some choices of interaction parameters; for other combinations, $\kappa_{xy}(T)$ has a local minimum/maximum and can even change sign with the \textcolor{black}{temperature increase} \cite{lieb,kagome_mook2014}. This heterogeneous character of the $\kappa_{xy}(T)$ for different parameters may be related to different topological phases of the insulating system, indexed by the Chern numbers of the bands. As mentioned above, our system is an insulator with a single trivial phase (the Chern number is zero for any combination of parameters), so we wouldn't expect any change in the character of the curve for different parameters. The same can be said about the other transport coefficients. \textcolor{black}{The only} change we detected in the transport coefficients was from a quantitative nature (see Figure \ref{2D} for an example).

\begin{figure}[h!]
\centering
\includegraphics[width=0.5\textwidth]{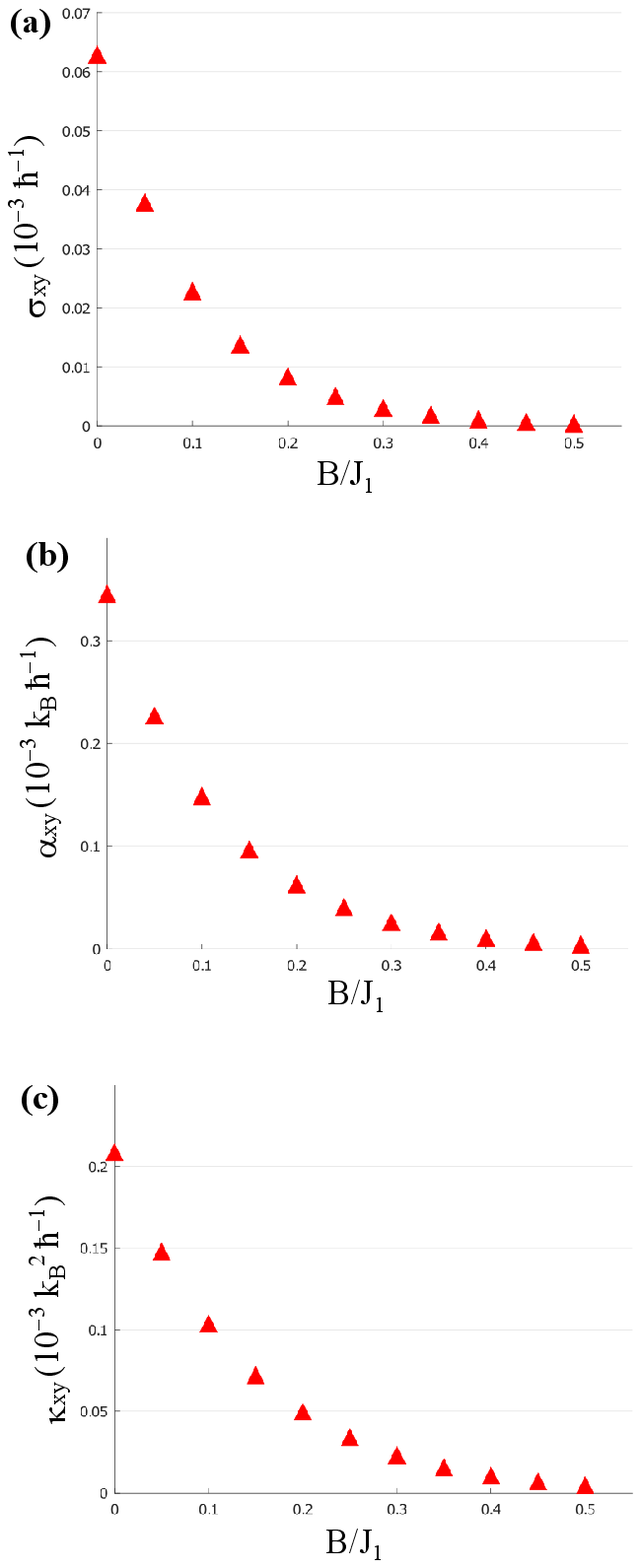}
\caption{(a) Spin Hall conductivity, (b) spin Nernst coefficient and (c) thermal Hall conductivity versus magnetic field for $S=1/2$, $T=0.1$, $J_1=1$, $J_2=0.5$, $J_3=0$, $\lambda=1.2$ and $D=0.2$.}
\label{fig9}
\end{figure}

\begin{figure}[h!]
\centering
\includegraphics[width=0.5\textwidth]{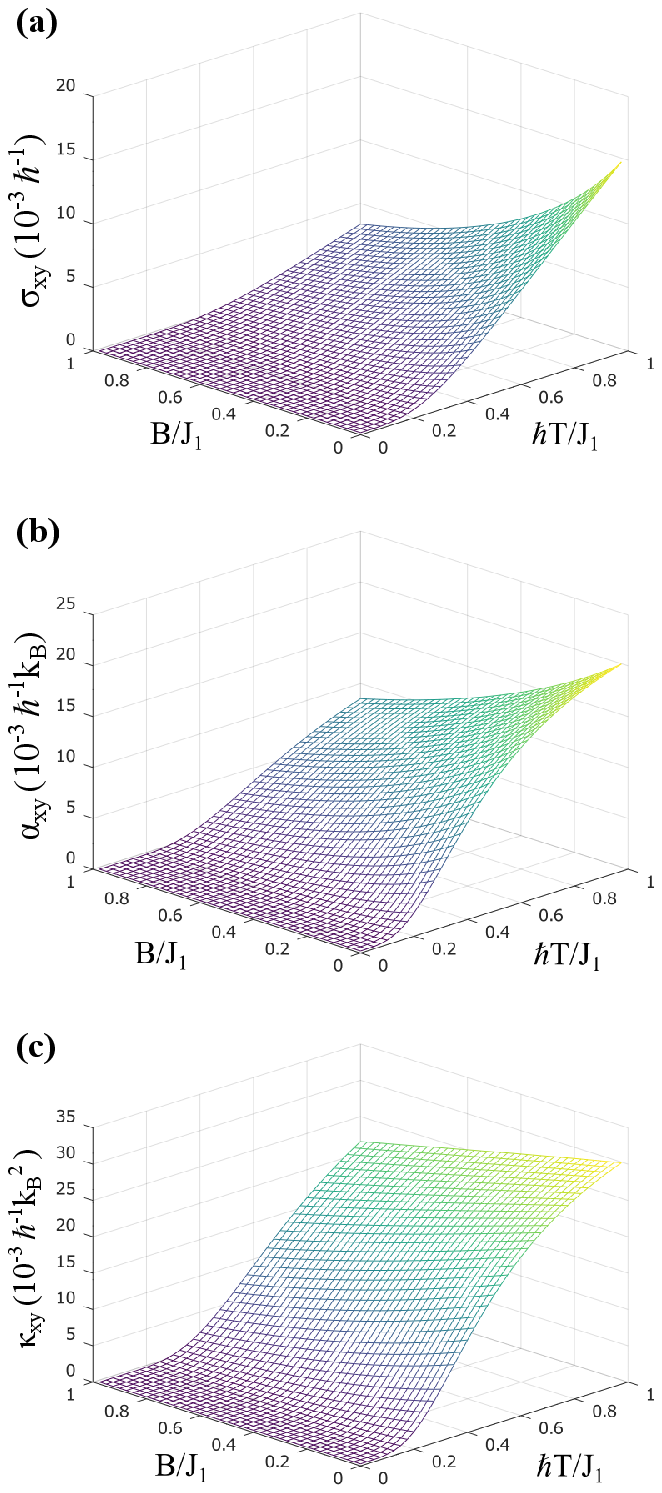}
\caption{(a) Spin Hall conductivity, (b) spin Nernst coefficient and (c) thermal Hall conductivity plotted against temperature and magnetic field. The theory parameters are $S=1/2$. $J_1=1$, $J_2=0.5$, $J_3=0$, $D=0.1$, $\lambda=1.2$.}
\label{fig10}
\end{figure}

\textcolor{black}{Figure \ref{fig9} shows the three transport coefficients as functions of the applied perpendicular magnetic field $B$ for fixed temperature. From Eqs. (\ref{eqdispersion}) and (\ref{berry_eq}), we see that a magnetic field does not affect the Berry curvature but increases $\omega_+$ and $\omega_-$. So, for a given $T$, a smaller number of magnons are excited in both bands. A strong magnetic field radically diminishes the thermal population difference between the bands, leading to a suppression of $\sigma_{xy}$, $\alpha_{xy}$ and $\kappa_{xy}$. This behavior was predicted for generic ferromagnetic 2D films in the dipolar regime \cite{prb89}, and was also observed in theoretical calculations on the checkerboard and Kagome lattices \cite{pires2020,LeeHanLee,kagomeAFM}. The coefficients' response to both temperature and the applied magnetic field can be seen in Figure \ref{fig10}}.

\begin{figure}[h!]
\centering
\includegraphics[width=1.0\textwidth]{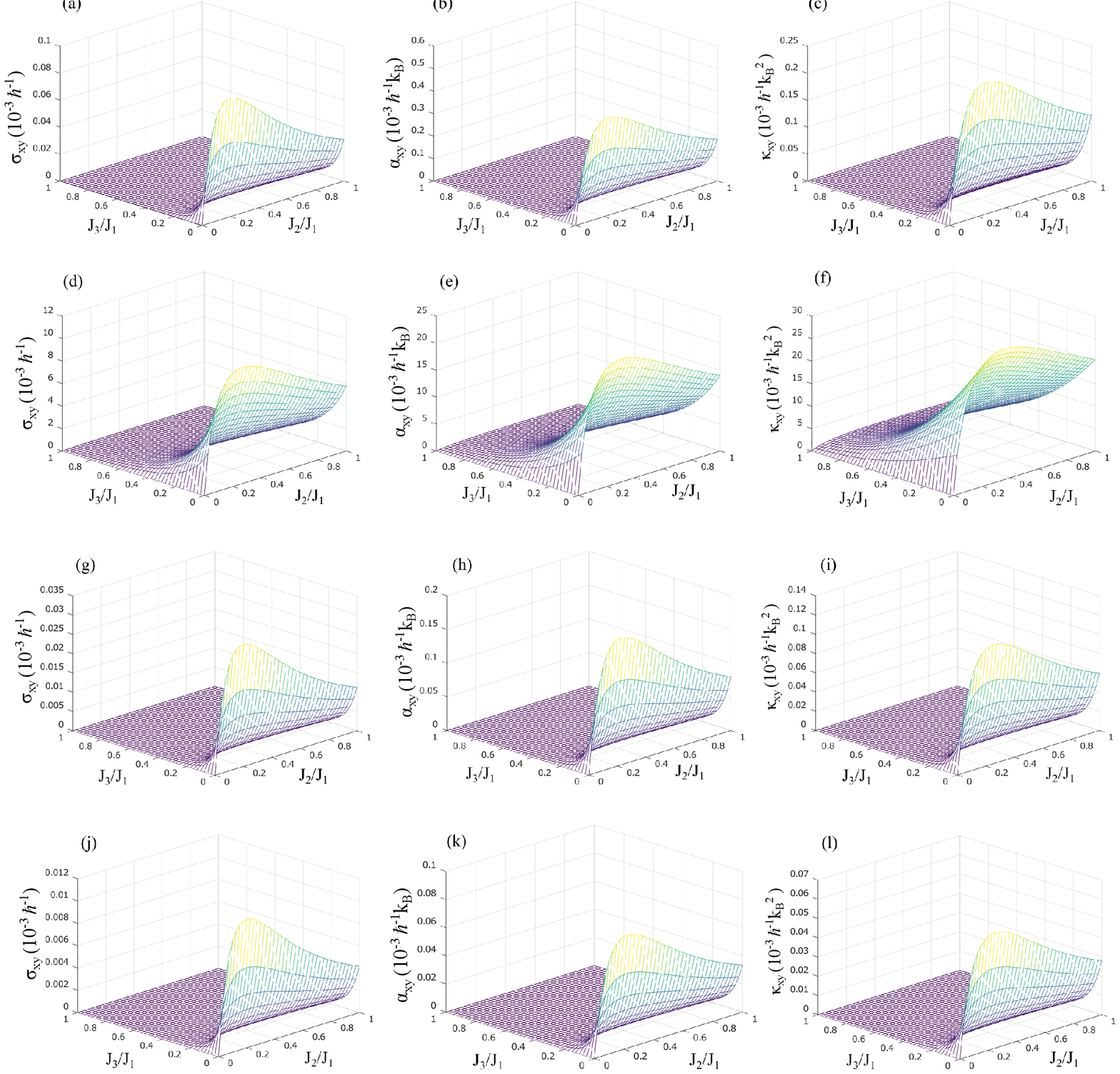}
\caption{All three transport coefficients as functions of the relative parameters $J_2/J_1$ and $J_3/J_1$ for different temperatures and applied magnetic fields: (a)-(c) $B=0$ and $T=0.1$. (d)-(f) $B=0$ and $T=0.5$. (g)-(i) $B=0.1$ and $T=0.1$. (j)-(l) $B=0.2$ and $T=0.1$. The other parameters are $S=1/2$, $J_1=1$, $J_2=0.5$, $J_3=0$, $D=0.1$ and $\lambda=1.2$.}
\label{fig11}
\end{figure}

\textcolor{black}{In Figure \ref{fig11}, we can see the dependence of the transport coefficients on the relative exchange parameters $J_2/J_1$ and $J_3/J_1$. An increase in $J_3$ leads to a decrease in the coefficients. On the other hand, all coefficients show a peak for a definite $J_2/J_1$ value. The exact $J_2$ value and peak height depend on the other parameters of the theory.}

\begin{figure}[h!]
\centering
\includegraphics[width=0.5\textwidth]{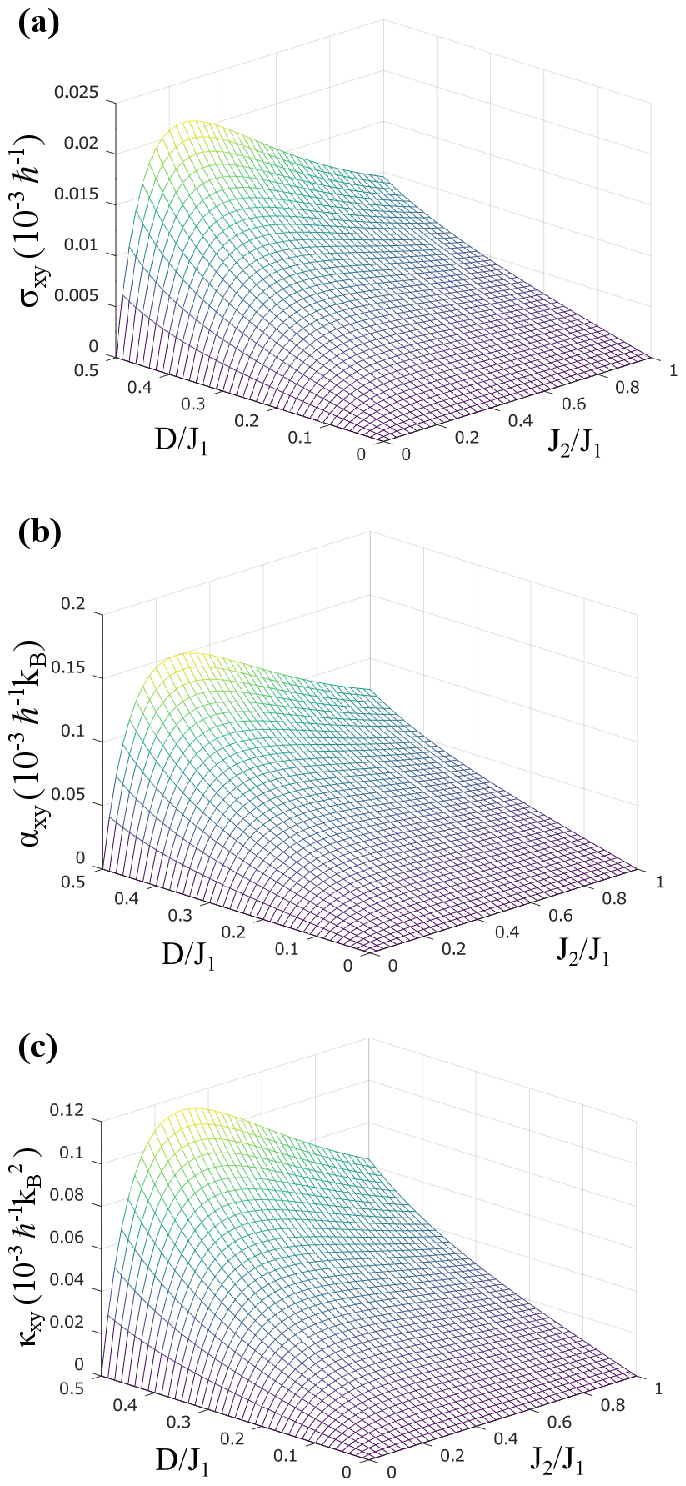}
\caption{(a) Spin Hall conductivity, (b) spin Nernst coefficient and (c) thermal Hall conductivity plotted against the relative parameters $J_2/J_1$ and $D/J_1$. The other theory parameters are $S=1/2$, $T=0.1$, $J_1=1$, $J_2=0.5$, $J_3=0$, $\lambda=1.2$, and $D=0.2$.}
\label{fig12}
\end{figure}

\textcolor{black}{In Figure \ref{fig12}, the coefficients were plotted against $J_2/J_1$ and $D/J_1$, and we can see again that they peak for a definite $J_2$ value. Concerning the DMI, the effect of increasing parameter $D$ is to raise all the transport coefficients. That can also be seen in Figures \ref{cond}-\ref{th}.}

\section{Conclusions}
\label{conclusion}

We studied a two-band ferromagnetic magnon model with the geometry of a modified Lieb lattice. An easy-axis anisotropy induces a gap in the $X'=(0,\pm \pi)$ point. When a Dzyaloshinskii-Moriya interaction is present between next-next-near \textcolor{black}{neighbors}, we find a non-vanishing Berry curvature due to time-reversal symmetry breaking.

The Chern numbers of the bands are null, so the insulating system has only one (trivial) phase. Nevertheless, the finite Berry curvature induces three magnon Hall-like effects whose transport coefficients were studied: the spin Hall effect, the thermal spin Hall effect, and \textcolor{black}{the} spin Nernst effect. The response of the coefficients to the temperature is monotonic without sign change, and resembles other topological magnon systems \textcolor{black}{in} the literature. The presence of an external off-plane magnetic field through a Zeeman interaction minimizes the transport effects, as we expected by thermodynamical considerations. 
\textcolor{black}{A strong exchange parameter $J_3$ also reduces the transport coefficients, while the DMI parameter tends to increase them. The transport coefficients are maximized for a definite $J_2/J_1$ value that depends on the other parameters of the theory.}

As far as we know, up to now there is no material described by the lattice studied here. Nevertheless, there is a variety of compounds described by the Lieb lattice. Thus, we believe that \textcolor{black}{by modifying the Lieb lattice, a compound could be synthesized where our model could be used.} Another possibility is in the field of optical lattices, where advances in synthesizing techniques make it possible to mimic DM interactions using \textcolor{black}{cold atoms trapped by laser beams \cite{Goldman2010,cold_atoms}}.

\section*{Declaration of competing interest}

The authors declare that they have no known competing financial interests or personal relationships that could have appeared to influence the work reported in this paper.

\section*{Acknowledgments}
This work was supported by CAPES (Coordenação de Aperfeiçoamento de Pessoal de Nível Superior) and CNPq (Conselho Nacional de Desenvolvimento Científico e Tecnológico).

\clearpage

\bibliographystyle{ieeetr}
\bibliography{refs}

\end{document}